\documentclass[11pt,aps,amsmath,latexsym,amsfonts]{JHEP3}

\usepackage{epsfig}
\usepackage{amsfonts,amssymb,amsmath}
\usepackage[hang]{subfigure}

\def\be{\begin{equation}}
\def\ee{\end{equation}}
\def\bea{\begin{eqnarray}}
\def\eea{\end{eqnarray}}
\def\nn{\nonumber}

\def\endignore{}
\def\ignore #1\endignore{} 

\def\Box{{\hbox{$\sqcup$}\llap{\hbox{$\sqcap$}}}}

\def\bd{\begin{displaymath}}
\def\ed{\end{diplaymath}}

\def\d{\mathrm{d}}
\def\m{\rm m}

\def\H{\mathcal{H}}

\def\nn{\nonumber}

\def\d{\mathrm{d}}

\def\({\left(}
\def\){\right)}

\def\bd{\boxdot}

\def\circa#1{\,\raise.3ex\hbox{$#1$\kern-.75em\lower1ex\hbox{$\sim$}}\,}
\makeatletter

\title{ Lifshitz solutions in  supergravity and string theory}

\author{
Ruth Gregory$^{a}$\!
\thanks{Email: r.a.w.gregory@durham.ac.uk}
 ,
Susha L.~Parameswaran$^b$
\thanks{Email: susha.parameswaran@fysast.uu.se}\ ,
 Gianmassimo Tasinato$^c$\!
\thanks{Email: g.tasinato@ThPhys.Uni-Heidelberg.de}\ ,
  Ivonne Zavala$^d$\!
\thanks{Email: zavala@th.physik.uni-bonn.de}
\\
$~^a$ Centre for Particle Theory,
Durham University, South Road, Durham, DH1 3LE, UK\\
$~^b$ Theoretical Physics, Uppsala University, Box 516, 
SE-751 20 Uppsala, Sweden \\
$~^c$ Institut f\"ur Theoretische Physik, Universit{\"a}t Heidelberg,
Philosophenweg  19, 69120 Heidelberg, Germany\\
$~^d$ Bethe Center for Theoretical Physics and
Physikalisches Institut der Universit\"at Bonn,
Nu\ss allee 12, D-53115 Bonn, Germany.}

\abstract{
We derive Lifshitz configurations in string theory for general dynamical
exponents $z\geq1$.  We begin by obtaining simple $Li \times \Omega$ solutions
to supergravities in diverse dimensions, with $\Omega$ a compact constant 
curvature manifold.  Then we uplift the solutions to ten dimensions,
providing configurations that correspond to warped compactifications 
in Type II string theory.
}
\preprint{DCPT-10/45\\ UUITP-32/10 }

\begin{document}

\section{Introduction}

In the past few years, the adS/CFT, \cite{Malda}, more generally 
called the gauge/gravity correspondence, has been explored to 
yield insights into strongly coupled field theory. While the
initial particle physics motivation may have been to glean a nonperturbative
understanding of QCD, a more unexpected but extremely fruitful 
dialogue has emerged with condensed matter physics (see e.g.\ \cite{CMP}), 
which also has many experimental systems believed to be governed by strongly
coupled physics. 

As suggested by the nomenclature adS/CFT,
the higher dimensional bulk gravitational theory asymptotically
tends to anti-de Sitter spacetime (adS), yielding a boundary theory which
is typically relativistically invariant. However, while many systems 
do display such scale invariance, there is also interest in 
systems displaying a more general dynamical scaling:
\be
t \to \lambda^z t \,, \;\; x \to \lambda x\,,
\label{lifdef}
\ee
where $z\neq1$ is the dynamical critical exponent. This 
splitting of space and time can seem counter-intuitive
for a relativistically invariant theory, however, it is always possible
to single out particular directions by choosing an appropriate
background. In \cite{kachru}, Kachru et al.\ showed how to
construct Lifshitz geometries by switching on fluxes with
nontrivial topological couplings. They constructed a four-dimensional
(4D) spacetime with two gauge fields (a 1- and 2-form) coupled via a 
topological ``$B_{2}\wedge F_{2}$" term. The model was, of course,
phenomenological, in that it was constructed to provide a holographic
dual with the requisite dynamical scaling properties, but it was hoped
that the model could be put on firmer footing by finding similar
Lifshitz geometries as solutions to String/M-theory. In this way, 
genuine dual quantum field theories could be constructed which would
hopefully encapsulate the physics of the condensed matter systems. 

Explicit Lifshitz solutions in string theory however have proven 
surprisingly difficult to find. The expectation was that standard 
techniques, such as flux compactifications, would be sufficient.  
Disappointingly, the first attempts with
reasonable and quite general Ans\"atze led only to various no-go
statements \cite{japanese, ulf}.  Several schemes which could support
Lifshitz geometries were discussed in \cite{sean}, and
significant progress was made with \cite{bala} and, most recently,
\cite{gauntlett}, where it was shown that Lifshitz solutions with 
the dynamical exponent $z=2$ can be embedded into
D=10 and D=11 supergravity, by compactifying on Einstein manifolds
fibred with a circle. These are important steps forward, however
they still are restricted to one particular value of the critical
exponent, and it would certainly be most interesting to 
see if they can be generalized to $z$
different from two, as well as to black hole backgrounds. 

In the present paper, we make a first step to answer these questions, and
provide a simple method to obtain explicit string constructions of
Lifshitz geometries with general dynamical exponents, $z \geq 1$  ($z=1$ being
the adS$_4$ limit).  Following a bottom-up approach, the idea 
is to start by looking for such solutions in $d$-dimensional supergravities,
and then to uplift them to ten dimensions.  This has been an often used
trick in the adS/CFT literature (see e.g.\  \cite{MN}).
Supergravities in diverse dimensions offer a host of simple 
adS$_q \times \Omega_{d-q}$ backgrounds, with $\Omega_{d-q}$ 
a constant curvature spherical, flat or hyperbolic space (see
\cite{salamsezgin} for a review).  Some of these, it
will be seen, can be straightforwardly generalized to simple $Li_{q}
\times \Omega_{d-q}$ geometries.

Our search led us to two examples of supergravity theories.  
First, we considered Romans gauged, massive, ${\cal N}=4$ 
supergravity in 6D, \cite{romans6d}, which has a solution
$Li_4 \times H_2$, with general $z \geq 1$ and $H_2$ a hyperboloid.  
The 2D hyperbolic space can then easily be rendered compact 
by modding out a non-compact
discrete subgroup of the isometry group.  This does not change the
local geometry, only the topology, leading to some genus $g$
Riemann surface, where $g$ depends on the choice of discrete
symmetry (for some useful references on compact hyperbolic spaces, see
\cite{MN}, \cite{kaloper}  and \cite{orlando}).  
Flux quantization on
the compact manifold leads to a topological restriction on  $z$ in
terms of the couplings of the 6D theory.  
The 6D solution can then be uplifted
to massive Type IIA supergravity following the analysis of
\cite{cvetic}, and the resulting configuration can be interpreted
in terms of intersecting D-branes of various dimensions.      
Second, we considered the gauged ${\cal N}=4$ supergravity in 5D, 
also by Romans \cite{romans5d}. This has solutions $Li_3 \times H_2$, 
again with general $z \geq 1$ 
(up to quantization conditions). They can be uplifted to Type
IIB supergravity using the results of \cite{lpt}, which
leads to configurations that
can be interpreted in terms of intersecting D3 branes, or further to
11D supergravity using \cite{5dto11d, 5dto11dgen}.

The paper is organized as follows.  In Section \ref{sect6d} we review the
6D supergravity theory and present 4D Lifshitz solutions, and then uplift
them to massive Type IIA supergravity. Then, in Section
\ref{sect5d}, we present an analogous discussion for 3D Lifshitz
solutions in 5D supergravity, uplifting them
to Type IIB string theory.  Finally, we
conclude in Section \ref{sectdisc}.  Throughout the paper, our
signature is mostly minus, and the curvature tensors are defined such
that the scalar curvature of a sphere is negative. Moreover we choose 
conventions such that the $d$-dimensional Newton constant
is $\kappa_d^2 = 2$.

\section{$Li_4$ solutions in massive Type IIA via 6D}
\label{sect6d}

The first Lifshitz solution we present has four infinite (spacetime)
dimensions, and arises from a compactification of ${\cal N}=4$, 6D
gauged massive supergravity. It can provide a dual description for a
$2+1$-dimensional strongly coupled field theory. We first review the
action and equations of motion of Romans, \cite{romans6d}, then show
that a simple $Li_4 \times \Omega_2$ Ansatz reduces
the equations of motion to a set of algebraic equations that can
be solved straightforwardly.
The result is a $Li_4 \times H_2$ geometry, where
$H_2$ is a hyperbolic two dimensional geometry, which
can be taken to be  compact, (e.g.\ see \cite{MN}).
We demonstrate that the solution breaks supersymmetry, and show how to
uplift the configuration to massive Type IIA supergravity.  

The action for ${\cal N}=4$ 6D supergravity was worked out by Romans
in \cite{romans6d}, and we use the conventions found there. 
The bosonic field content consists of the metric, $g_{{\mu}{\nu}}$, 
dilaton, $\phi$, an anti-symmetric two-form gauge field, $B_{{\mu}{\nu}}$, 
and a set of gauge vectors, $(A_{\mu}^{(i)}, {\cal A}_{\mu})$, 
for the gauge group $SU(2)\times U(1)$. Fermions fill
out the supermultiplets, but they will not be of interest for our
purposes.  The bosonic part of the Lagrangian can be written as:
\bea
{\rm e}^{-1}\,{\cal L} &=& -\frac14 R
+\frac12 \partial^{\mu} \phi \partial_{\mu}
\phi-\frac{e^{-\sqrt{2} \phi}}{4}\,
\left( {\cal H}^{{\mu}{\nu}} {\cal H}_{{\mu}{\nu}}
+F^{(i)\,{\mu}{\nu}} F^{(i)}_{{\mu}{\nu}}\right) 
+\frac{e^{2\sqrt{2}\phi}}{12}
G_{{\mu}{\nu}{\rho}} G^{{\mu}{\nu}{\rho}} \nonumber \\
&&-\frac{{\rm e}}{8} \, \epsilon^{{\mu}{\nu}{\rho}{\lambda}{\sigma}{\tau}}\,
B_{{\mu}{\nu}} \times 
\left( {\cal F}_{{\rho}{\lambda}} {\cal F}_{{\sigma}{\tau}}
+{\rm m} B_{{\rho}{\lambda}} {\cal F}_{{\sigma}{\tau}}
+\frac{{\rm m}^2}{3} B_{{\rho}{\lambda}} B_{{\sigma}{\tau}}
+F^{(i)}_{{\rho}{\lambda}} F^{(i)}_{{\sigma}{\tau}} \right)\nonumber\\
&&+\frac18\left( g^2 e^{\sqrt{2} \phi}
+4 g {\rm m} e^{-\sqrt{2} \phi}-{\rm m}^2
 e^{-3 \sqrt{2} \phi}
\right) \, , \label{6dL}
\eea
where ${\rm e}$ is the determinant of the vielbein, the spacetime
indices ${\mu},{\nu},\dots$ run from $1,\dots, 6$, and the gauge indices,
$(i),(j),\dots$ run over $(1),(2),(3)$.  The field strengths are given by:
\bea
{\cal F}_{{\mu}{\nu}}&=&\partial_{\mu} 
{\cal A}_{\nu}-\partial_{\nu} {\cal A}_{\mu}\\
F^{(i)}_{{\mu}{\nu}}&=&\partial_{\mu} { A}^{(i)}_{\nu}-\partial_{\nu} 
{A}^{(i)}_{\mu}+g\, \epsilon^{ijk}{ A}^{(j)}_{\mu} { A}^{(k)}_{\nu}\\
G_{{\mu}{\nu}{\rho}}&=&3 \,\partial_{[{\mu}}\,B_{{\nu}{\rho}]} \, ,
\eea
and we also define:
\be
{\cal H}_{{\mu}{\nu}}={\cal F}_{{\mu}{\nu}}+{\rm m} B_{{\mu}{\nu}} \, .
\ee
Notice that the theory has two parameters: the gauge coupling, $g$, and
the mass parameter, $\m$.  Romans identified five distinct theories, 
labelled ${\cal N}=4^+$ (for $g>0, \m>0$), ${\cal N}=4^-$ (for $g<0, \m>0$), 
${\cal N}=4^g$ (for $g>0, \m=0$), ${\cal N}=4^{\m}$ (for $g=0, \m>0$) 
and finally ${\cal N}=4^0$ (for $g=0,\m=0$).  
 
The equations of motion that follow from the Lagrangian (\ref{6dL}) read:
\bea
&&R_{{\mu}{\nu}} = 2 \partial_{\mu} \phi \partial_{\nu} \phi 
+ g_{{\mu}{\nu}} P(\phi) +
e^{2\sqrt{2}\phi} \left(G_{\mu}^{\,\,\,{\rho}{\lambda}} 
G_{{\nu}{\rho}{\lambda}} - g_{{\mu}{\nu}}
G^{{\rho}{\lambda}{\sigma}} G_{{\rho}{\lambda}{\sigma}} 
\right) \nonumber \\
&& \hskip 1cm
- e^{-\sqrt{2}\phi} \left(2 \H_{\mu}^{\,\,{\rho}} \H_{{\nu}{\rho}} 
+ 2 F_{\mu}^{\,\,{\rho}\,(i)} F_{{\nu}{\rho}}^{(i)} 
- \frac14 g_{{\mu}{\nu}} \left(\H^{{\rho}{\lambda}}\H_{{\rho}{\lambda}} 
+ F^{{\rho}{\lambda}\,(i)}F_{{\rho}{\lambda}}^{(i)} \right)
\right) \label{6deom1}  \\
&&\Box \phi = \frac{\partial P}{\partial\phi} + \frac13
\sqrt{\frac{1}{2}} e^{2\sqrt{2}\phi} G^{{\mu}{\nu}{\rho}} 
G_{{\mu}{\nu}{\rho}} + \frac12
\sqrt{\frac12} e^{-{\sqrt{2}\phi}} \left(\H^{{\mu}{\nu}} \H_{{\mu}{\nu}} 
+ F^{{\mu}{\nu}\,(i)} F_{{\mu}{\nu}}^{(i)} \right)   \\
&&D_{\nu} \left( e^{-\sqrt{2}\phi} \H^{{\nu}{\mu}} \right) 
= \frac16 {\rm e} \,
\epsilon^{{\mu}{\nu}{\rho}{\lambda}{\sigma}{\tau}} \H_{{\nu}{\rho}} 
G_{{\lambda}{\sigma}{\tau}} \\
&&D_{\nu} \left( e^{-\sqrt{2}\phi} F^{{\nu}{\mu}\,(i)} \right) 
= \frac16 {\rm e} \, 
\epsilon^{{\mu}{\nu}{\rho}{\lambda}{\sigma}{\tau}} 
F_{{\nu}{\rho}}^{(i)} G_{{\lambda}{\sigma}{\tau}} \\
&&D_{\rho} \left(e^{2\sqrt{2}\phi} G^{{\rho}{\mu}{\nu}} \right) 
= - {\rm m}
e^{-\sqrt{2}\phi} \H^{{\mu}{\nu}} - \frac14 {\rm e} \, 
\epsilon^{{\mu}{\nu}{\rho}{\lambda}{\sigma}{\tau}} \left(\H_{{\rho}{\lambda}}
\H_{{\sigma}{\tau}} + F_{{\rho}{\lambda}}^{(i)} F_{{\sigma}{\tau}}^{(i)}
 \right) \,, 
\label{6deom5}
\eea
where we have defined the scalar potential function:
\be
P(\phi)=\frac18\left( g^2 e^{\sqrt{2} \phi}
+4 g {\rm m} e^{-\sqrt{2} \phi}-{\rm m}^2
 e^{-3 \sqrt{2} \phi}
\right) \,.
\ee

In order to solve these equations, we make a simple Ansatz for 
the solution: that the metric consists of a direct product 
between a 4D Lifshitz geometry, and a constant 
curvature 2D internal space:
\be
\label{4DAns}
d s^2\,=\,L^2\left( 
r^{2 z} d t^2-r^2 d x_1^2 -r^2 d x_2^2-\frac{d r^2}{r^2}
\right) - d\Omega_2^2 \,,
\ee
where the `internal' 2D part of the metric takes the form 
of flat
space, or:
\be
d\Omega_2^2 = a^2(d\theta^2 + \sin^2\theta \d\varphi^2) \qquad {\rm or} \qquad
d\Omega_2^2 = \frac{a^2}{y_2^2}\left(d y_1^2+ d y_2^2 \right)\, ,
\label{twodgeom}
\ee
with $a$ the radius of curvature of the sphere or hyperboloid
respectively. In (\ref{4DAns}). the parameter $z$ is the dynamical exponent, 
which measures the anisotropic scaling symmetry:
\be
t \rightarrow \lambda^{z} t, \qquad x^i \rightarrow \lambda x^i, \qquad r
\rightarrow \lambda^{-1} r\, .
\ee
Gauge field backgrounds that are invariant under the chosen
symmetries are: 
\bea
F^{(3)}_{tr}\,&=&\, \alpha L^2 r^{z-1}\hskip 0.5cm,\hskip 0.5cm 
F^{(3)}_{y_1y_2}= \gamma \, {\rm e}_2  \\
G_{x_1 x_2 r}&=& \beta L^3 r \hskip 0.5cm \Rightarrow \hskip 0.5cm
B_{x_1 x_2}\,=\,\frac{\beta}{2} L^3 r^2 \, ,
\eea
with ${\rm e}_2$ the determinant of the zweibein on $\Omega_2$,
and the scalar field is constant with value $\phi_0$.  

We take $z$ to be some fixed number in the solution. When
$z=1$ the Ansatz reduces to the standard adS$_4 \times \Omega_2$ one.
These solutions were discussed in Romans'  paper
\cite{romans6d}, where they were found to exist in the ${\cal N}=4^+$
theory, with the internal space being $H_2$.  For a special value of the
parameters\footnote{Note, however, that Romans' condition on the
  couplings $g=2{\m}$ 
  can be relaxed by allowing for a non-zero constant dilaton.} these 
solutions are supersymmetric thanks to an embedding of the spin
connection in the gauge connection, with half of the original supersymmetries
surviving. 
Later, in \cite{Nunez:2001pt}, these solutions and their
deformations were studied
 in the context of adS/CFT, along with similar adS$_4 \times
S^2$ geometries.  Beware, however, that the fixed point $S^2$
configurations are not solutions to the supersymmetry conditions 
nor the equations of
motion\footnote{One of the components of the
gravitino supersymmetry equation (component $\mu=6$ in the case of  
an internal sphere, and component $\mu=5$ for the hyperboloid) gives
rise to an extra supersymmetry condition, $a g = 1$, not explicitly
written in
\cite{Nunez:2001pt}.  Together with the
BPS conditions given in \cite{Nunez:2001pt}, this singles out $H_2$ as the only
supersymmetric fixed-point solution.  It is also 
straightforward to show that the $S^2$ geometry does not solve the 
second order Einstein equations.  We thank the authors of
\cite{Nunez:2001pt} for discussions on this point.}.
Therefore, in the following we focus on the hyperbolic
internal manifold in the ${\cal N}=4^+$ theory. 

In the familiar case of an adS$_4\times H_2$ Ansatz, the equations of
motion (\ref{6deom1}$-$\ref{6deom5}) reduce to a set of algebraic
equations.  The same is true for the more general Lifshitz Ansatz.  For
convenience, we make the following rescalings,
which absorb the two quantities $\phi_0$ and $L$ into
a redefinition of the other parameters:
\bea
\hat\alpha &=& L \alpha e^{-\phi_0/\sqrt{2}}  \hskip1cm   
\hat\beta = L \beta e^{\sqrt{2}\phi_0}  \hskip1cm   
\hat\gamma = L \gamma e^{-\phi_0/\sqrt{2}}  \\
\hat g &=& L g e^{\phi_0/\sqrt{2}} \hskip14mm
\hat a = a/L \hskip16mm
\hat{\m}= L\, {\m}\, e^{-3\phi_0/\sqrt{2}} \,.
\eea
In essence, we are interchanging the freedom to choose the parameters
$\phi_0$ and $L$ with the freedom to choose the Lagrangian
parameters $\hat g$ and $\hat {\m}$. It is of course straightforward to 
recover the values of $\phi_0$ and $L$ by inverting these relations. 
With our Ansatz and these redefinitions, the field equations 
(\ref{6deom1}$-$\ref{6deom5}) simplify to the following set 
of algebraic relations.
\bea
\,z\,\hat \beta&=&\frac{{\rm \hat m}^2}{2}\, 
\,\hat \beta +2 \hat \alpha \hat \gamma
\label{6d3f} \\
\hat  \alpha&=&\hat \gamma  \hat \beta  \label{6d2f} \\
0&=&\frac14 \left(\hat{g}^2 -4 \hat g\,  \hat {\rm m} 
+3 \hat{{\rm m}}^2 \right)-2 \hat \beta^2 
+\left( \frac{\hat {\rm m}^2 \hat \beta^2 }{4}
- \hat \alpha^2+ \hat \gamma^2\right) 
\label{6ddil} \\ 
z (2+z)&=&{\cal P}+\hat  \beta^2 
+\left( \frac{\hat {\rm m}^2 \hat \beta^2}{8}
+\frac{3 \hat \alpha^2}{2}+\frac{\hat \gamma^2}{2}\right) 
\label{6dtt}\\
2+z&=&{\cal P}- \hat \beta^2 
+ \left( -\frac{3 \hat {\rm m}^2 \hat \beta^2}{8}
-\frac{ \hat \alpha^2}{2}+\frac{\hat \gamma^2}{2}\right) 
\label{6dxx}\\
2+z^2&=&{\cal P}- \hat \beta^2 
+ \left( \frac{ \hat {\rm m}^2 \hat \beta^2}{8}
+\frac{ 3 \hat \alpha^2}{2}+\frac{\hat \gamma^2}{2}\right) 
\label{6drr}\\
\frac{ 1}{\hat a^2}&=&{\cal P}+ \hat \beta^2
+ \left( \frac{\hat  {\rm m}^2 \hat \beta^2}{8}
-\frac{\hat\alpha^2}{2}-\frac{3 \hat \gamma^2}{2}\right) 
\label{6d55}
\eea
where
\be
{\cal P}=\frac18 \left(\hat g^2  +4 \hat g\, \hat {\rm m} 
- \hat {\rm m}^2 \right) \,.
\ee
Although this would appear to be a system of seven equations in
six independent variables, there is a Bianchi identity which
relates a combination of (\ref{6d3f}) and (\ref{6d2f}) to a
combination of (\ref{6dtt}), (\ref{6dxx}) and (\ref{6drr}).
Thus a solution exists also for $z \neq 1$, and is given by: 
\bea
\hat \beta^2 &=& z-1\\
\hat \alpha^2 &=& \hat \gamma^2 (z-1)\\
\hat \gamma^2&=&\frac{ (2+z) (z-3)\pm 2 \sqrt{2(z+4)}}{2 z} 
\label{exphatc}  \\
\hat g^2&=& 2 z(4+z)\\
\frac{\hat{m}^2}{2}&=&\frac{6+z \mp 2\sqrt{2(z+4)}}{z}\\
\frac{1}{\hat a^2} &=& 6+3z \mp 2\sqrt{2(z+4)} \, .
\eea

The internal hyperbolic space can be taken to be non-compact or
compact.  For our purposes, the compact case is most interesting, and
we obtain this case as follows.  The 2D hyperboloid can be written as a coset
space $SO(1,2)/SO(2)$, and modding out by a freely acting, discrete
non-compact subgroup of the isometry group, $SO(1,2)$, we arrive at a
compact manifold.  This manifold can be seen as a Riemann surface of
some genus, which depends on the choice of subgroup (see e.g.\ 
\cite{orlando}). Such spaces have been much discussed
in the compactification literature, beginning with \cite{kaloper}.
Notice now that, although classically solutions exist for all dynamical
exponents $z \geq 1$, the quantization condition for the
flux threading the 
compact internal manifold provides a relation between $z$ and the parameters of the 6D theory, $g, {\m}$.

It is easy to see that all the Lifshitz solutions ($z \neq 1$) break
supersymmetry. The relevant supersymmetry transformations are the
fermionic gravitini and dilatini ones. They take the form \cite{romans6d}:
\bea
&&\delta \psi_{{\mu}\, a} = D_{{\mu}} \epsilon_a -\frac{1}{8\sqrt{2}} 
\left(g e^{\phi/\sqrt{2}} + \m \,e^{-3\phi/\sqrt{2}}\right) 
\gamma_{\mu} \gamma_7 \epsilon_a 
- \frac{1}{24} e^{\sqrt{2} \phi} \gamma_7 \gamma^{{\rho}{\lambda}{\sigma}}
G_{{\rho}{\lambda}{\sigma}} \gamma_{\mu} \epsilon_a \nonumber \\
&&\qquad\quad - \frac{1}{4\sqrt{2}} e^{-\phi/\sqrt{2}}
\left(\gamma_{\mu}^{{\nu}{\rho}} - 6 \delta_{\mu}^{\nu} \gamma^{\rho} \right) 
\left(\frac12 \H_{{\mu}{\nu}} \delta_a^b 
+ \gamma_7 F_{{\mu}{\nu}}^{(i)} T_a^{(i)\,\,b} \right) \epsilon_b\\
&& \delta \chi_a = \frac{1}{\sqrt{2}} \gamma^{\mu} (\partial_{\mu} \phi)
\epsilon_a + \frac{1}{4\sqrt{2}}\left(g e^{\phi/\sqrt{2}} - 3 \m\,
e^{-3\phi/\sqrt{2}} \right) \gamma_7 \epsilon_a - \frac{1}{12}
e^{\sqrt{2} \phi} \gamma_7 \gamma^{{\rho}{\lambda}{\sigma}} 
G_{{\rho}{\lambda}{\sigma}} \gamma_{\mu} \epsilon_a \nonumber \\
&& \qquad\quad \frac{1}{2\sqrt{2}} \gamma^{{\mu}{\nu}} \left(\frac12
\H_{{\mu}{\nu}} \delta_a^b + \gamma_7 F_{{\mu}{\nu}}^{(i)} T_a^{(i)\,\,b} \right)
\epsilon_b \, .
\eea
It is sufficient to consider the dilatini transformation.  
In this case, we can see that the non-trivial fluxes that 
support the Lifshitz solution would require us to impose 
four independent projection conditions for $\delta
\chi_a = 0$, and so we cannot preserve supersymmetry.  
This fact can also be seen at the level of the ten dimensional, 
uplifted solution, as we show now.
 
Six dimensional  Romans' gauge supergravity can be uplifted via 
an $S^4$ to massive Type IIA supergravity in ten dimensions \cite{cvetic}.
Indeed, using the results of \cite{cvetic}, it is trivial to uplift any
solution of the 6D field equations to a solution of the 10D
equations of motion.  We now do so for the $Li_4 \times H_2$
configuration identified above.
In order to go  from the six dimensional  action used in \cite{cvetic}
to Romans' conventions used here, we make  the following 
redefinitions\footnote{In addition, we take conventions 
where $\kappa_6^2 =2$ rather than $\kappa_6^2= 1/2$ as was 
taken in \cite{cvetic}. } 
(a tilde denotes quantities in \cite{cvetic} notation): 
\bea
&& \widetilde g_{\mu\nu}  = - g_{\mu\nu} \,, \qquad \qquad \quad \,\,\,\,
\widetilde \phi - 2\widetilde \phi_0 =   -2\phi\,,\\  
&& e^{2\sqrt{2}\,\widetilde \phi_0} = \frac{3\m}{g} \,, \qquad \qquad \qquad 
\,{ \widetilde g} = \frac{\left(3{\rm m}  g^3\right)^{1/4} }{2}\,,\\  
&& \frac12 e^{-\sqrt{2}\, \widetilde \phi_0} 
\widetilde B_{2}  = B_2\,,  \qquad \qquad  
\frac12 e^{\sqrt{2}/2\, \widetilde \phi_0} \widetilde F^{(i)}_2  = F_2^{(i)} \,.
\eea
Using this dictionary, we can now write the 10D solution using
the formulae in \cite{cvetic}. Defining 
\bea 
&& k_0 = e^{\phi_0/\sqrt{2}}\left(\frac{g}{3\m}\right)^{1/4}  \\
&& C(\rho)= \cos \rho \,, \qquad S(\rho)= \sin \rho \\
&& \Delta(\rho) = k_0 \, C^2 + k_0^{-3}\,S^2  \\ 
&&  U(\rho) = k_0^{-6}\,S^2 - 3 k_0^2\,C^2 + 4 k_0^{-2}\,C^2 -6
k_0^{-2}  \, ,
\eea
as well as the constants
\bea
&& k_1 = \frac{8}{g^2} \frac{g}{3\m} e^{\sqrt{2}\, \phi_0}, \quad k_2 =
\frac{2}{g^2} \left(\frac{g}{3\m}\right)^{1/4} e^{-\phi_0/\sqrt{2}},
\quad k_3 = -\frac{4\sqrt{2}}{3}\frac{1}{g^3}
\left(\frac{g}{3\m}\right)^{3/4}, \quad     \\ \nonumber \\
&& k_4 = 3 g^2 e^{2\sqrt{2}\phi_0} k_3 \;,
\quad k_5 =  3 g k_3 \;, 
\quad k_6 = -  \frac{2\sqrt{2}\,e^{-3\phi_0/\sqrt{2}}}{g^2}, 
\quad k_7 = 2 \left(\frac{3\m}{g}\right)^{1/2} \,,\nonumber
\eea
the ten dimensional, uplifted configuration that results is
\bea
&& ds_{10}^2  = S^{1/12} \,k_0^{1/8}\left[  \Delta^{3/8}(Li_4\times {
H}_2) - k_1 \Delta^{3/8} \,d\rho ^2 - k_2 \Delta^{-5/8} C^2
\sum_i^3 (h^{(i)})^2 \right] \, ,
\nonumber \\
&& {\bf F_4} = k_3 \, S^{1/3} \, C^3  \, \Delta^{-2} \,U \, 
d\rho  \wedge  \epsilon_3 + k_4 \,S^{1/3}C \, \star_6 G_3 \wedge d\rho\nonumber \\ 
&&\qquad+ k_5  \, S^{1/3} \,  C \, F_2^{(3)} \wedge h^{(3)}\wedge d\rho
+ k_6 \, S^{4/3} \, C^2 \Delta^{-1} \,F_2^{(3)} \wedge
\sigma^{(1)}\wedge \sigma^{(2)} \, ,\label{ulift1} \\
&& {\bf G_3} = k_7  \, S^{2/3}  \, G_3  \, , \quad {\bf F_2} = 0 \, ,
\nonumber \\
&&  e^{\Phi} = S^{-5/6} \,  \Delta^{1/4} \,  k^{-5/4}_0 \,,\nonumber
\eea
where 
\be
h^{(i)}  = \sigma^{(i)} - g \, A_1^{(i)}\,,
\ee
with $\sigma^{(i)} $ the left-invariant 1-forms on $S^3$, 
and $\epsilon_3 =h^{(1)}\wedge h^{(2)} \wedge h^{(3)}$.   
The parameters of the 6D theory are related to the Type
IIA mass parameter via ${\bf m} = 
\left(2 \,{\m} \, g^3/27\right)^{1/4}$. 
Notice that the ten dimensional RR ${\bf F_2}$ field strength vanishes,
while the  RR  ${\bf F_{4}}$ field and the NS ${\bf G_3}$ field are
switched on.  The uplifted solution (\ref{ulift1}) contains the six dimensional
fields  $F_2^{(3)}$, $G_3$ and $\star_6 G_3$, which we recall here:
\bea\label{sdf2up}
F_2^{(3)}  &=& \hat \gamma\, e^{\frac{\sqrt{2} \phi_0}{2}}  
\,\left[\sqrt{z-1} L\, r^{z-1} dt \wedge dr  + \frac{a^2}{L\,y_2^2}  \, 
dy_1\wedge dy_2 \right] \nonumber \\
G_3 &=& e^{-\sqrt{2}\phi_0} \, L^2  \, \sqrt{z-1}\, r\, dx_1\wedge dx_2
\wedge dr \\
\star_6 G_3 &=&   e^{-\sqrt{2}\phi_0} \, \frac{a^2}{y_2^2} 
\, \sqrt{z-1}\, r^{z}\, dt \wedge dy_1 \wedge dy_2  \, .\nonumber
\eea
We can see the effects of the various charges in the 10D
solution. The 3-form flux, $G$, lifts both directly to the 3-form
${\bf G}_3$, as well as contributing to the 4-form. ${\bf F}_4$
also contains a geometric term and contributions from the
gauge field $F_2^{(3)}$, which now appears as a Kaluza-Klein (KK) field in
the angular directions of the transverse 4D space\footnote{The
presence of a KK gauge field is similar to the Type IIB Lifshitz solution
presented in \cite{gauntlett}.}.  
The fluxes in 10D will give rise to
further quantization conditions.   At $\rho = \pi/2$, the metric has a
coordinate  
singularity, as can be seen by considering the limit
 $k_2/\Delta k_1\to 
1/4$ as  $ \rho \to \pi/2$. 
Due to the overall factors $S^{1/12}$, 
the metric is also singular at $\rho=0,\pi$.

The previous configuration can be  interpreted as a system of  D4-D8 branes, 
intersecting with D2 branes and an NS5 brane. We can understand this as
follows. Take first $z=1$, in this case the NS gauge field
turns off, as well as  the parts of the RR ${\bf F_{4}}$ field 
that contain a time component.  Therefore, ${\bf F_{4}}$ is
magnetically sourced by D4 branes, whereas ${\bf F_{10}}$ (associated
with the IIA mass) is sourced electrically by D8 branes.  One ends up
with a D4-D8 system, which can preserve supersymmetry for certain
configurations, see e.g.\  \cite{Nunez:2001pt}.  In contrast, when $z>1$,  the NS field, as well as the 
remaining  components of the RR ${\bf F_{4}}$ field are turned on.   These
fields are, respectively, sourced magnetically by NS5 branes and
electrically by D2 branes, which all
intersect the previous D4-D8 system.  Consequently, the four dimensional 
space-time symmetry is reduced from Lorentz to Lifshitz, and
supersymmetry is broken completely.

\section{$Li_3$ solutions in Type IIB via 5D}
\label{sect5d}

In this section we present three dimensional Lifshitz string solutions,
which are dual to field theories in $1+1$ dimensions.  We find
the solutions via compactifications of ${\cal N}=4$ 5D gauged supergravity,
which can be uplifted to Type IIB supergravity in ten dimensions, or
eleven dimensional supergravity.  The steps are very similar to the 6D
case of the previous section.  After presenting the 5D action and
field equations, we find the general $Li_3 \times \Omega_2$ solutions,
where $\Omega_2$  again turns out to be restricted to a hyperboloid.
We show that supersymmetry is broken, and then we
explain how to embed the solution in Type IIB string theory.

The ${\cal N}=4$ 5D gauged supergravity was developed in \cite{romans5d}, and
our conventions are the same as the original reference.  The
field content consists of the metric, $g_{\mu\nu}$, dilaton, $\phi$,
gauge fields, 
$(A_{\mu}^{(i)}, {\cal A}_{\mu})$, for an $SU(2) \times U(1)$ gauge group, two
antisymmetric tensor fields, $B_{\mu\nu}^{\alpha}$ ($\alpha$ indicates
the real two dimensional vector representation of $U(1)$),  and the fermionic
partners.  The bosonic part of the Lagrangian is:
\bea
{\rm e}^{-1}{\mathcal L}&=&
-\frac14 R+\frac12 D_\mu \phi D^\mu \phi-\frac14 \xi^{-4} 
{\cal F}^{\mu\nu} {\cal F}_{\mu\nu}-\frac14 \xi^2 \left(   
F_{\mu\nu}^{(i)} F^{\mu\nu\,\,(i)} + B^{\mu\nu\alpha} B_{\mu\nu}^{\alpha}
\right) \nonumber \\
&& \quad +\frac14\, 
 {\rm e} \,\epsilon^{\mu\nu\rho\sigma\lambda} \left(\frac{1}{g_1}
 \epsilon_{\alpha\beta} B_{\mu\nu}^{\alpha} D_{\rho} B_{\sigma\lambda}^{\beta}-
\,F_{\mu\nu}^{(i)} F_{\rho\sigma}^{(i)}  {\cal A}_\lambda \right) + P(\phi) \, , \label{5dL}
\eea
where we have defined $\xi \equiv e^{\sqrt{\frac23}\phi}$, and the
scalar field potential is 
\be
P(\phi)\,=\,\frac{g_2}{8}\,\left(g_2 \,\xi^{-2}+2\sqrt{2}\,g_1 \,\xi\right) \, .
\ee
Also, the field strengths are as usual:
\bea
{\cal F}_{\mu\nu} &=& \partial_{\mu} {\cal A}_{\nu} - \partial_{\nu}
{\cal A}_{\mu} \\
F_{\mu\nu}^{(i)} &=&  \partial_{\mu} A_{\nu}^{(i)} - \partial_{\nu} 
A_{\mu}^{(i)} +
g_2 \,\epsilon^{ijk} A_{\mu}^{(j)} A_{\nu}^{(k)} \label{5dnafs}\, ,
\eea
and ${g_2}, {g_1}$ 
are the gauge couplings for $SU(2) \times U(1)$, respectively.
The 5D gauged supergravity thus has two independent
parameters, ${g_1}, {g_2}$, which give rise to three
physically distinct theories.  Following Romans \cite{romans5d},
 when ${g_1}
{g_2} >0$, we call the theory ${\cal N}=4^+$, when ${g_2} = 0$ we
call it ${\cal N}=4^0$, and when ${g_1} {g_2} <0$ we have ${\cal N}=4^-$.

The equations of motion that result from the Lagrangian (\ref{5dL}):
\bea
R_{\mu\nu}&=&2\partial_\mu \phi \partial_\nu \phi
+\frac43 g_{\mu\nu}P-\xi^{-4}\left(
2 {\cal F}_{\mu\rho} {\cal F}_\nu^\rho
-\frac13 g_{\mu\nu} {\cal F}_{\rho\sigma}{\cal F}^{\rho\sigma}
\right)\nonumber\\
&&-\xi^2\left( 
2 F_{\mu\rho}^{(i)} F_{\nu}^{\rho\,\,(i)}-\frac13 g_{\mu\nu}\,
F_{\rho \sigma}^{(i)} F^{\rho \sigma\,\,(i)}
\right)\\
\Box \phi &=&\frac{\partial P}{\partial \phi}+
\sqrt{\frac{2}{3}}{\cal F}_{\mu\nu} {\cal F}^{\mu\nu}-\sqrt{\frac{1}{6}}
\xi^2 \,F_{\rho \sigma}^{(i)} F^{\rho \sigma\,\,(i)}\\
D_\nu\left(
\xi^{-4} {\cal F}^{\nu\mu}
\right)&=&\frac14 e^{-1} \epsilon^{\mu\nu\rho\sigma \tau}
\,F_{\nu\rho}^{(i)} F_{\sigma \tau}^{(i)}\\
D_\nu\left( \xi^2 F^{\nu \mu\,(i)}\right)&=&
\frac12 e^{-1} \epsilon^{\mu\nu\rho\sigma \tau}
\,F_{\nu\rho}^{(i)} {\cal F}_{\sigma \tau} \, ,
\eea
where we have set the antisymmetric tensor fields
$B_{\mu\nu}^{\alpha}$ to zero, in view of our Ansatz below.  The
maximally symmetric solutions to these equations were studied in 
\cite{romans5d}.  The ${\cal N}=4^+$ theory was found to have vacua of the
form adS$_3 \times \Omega_2$ for $\Omega_2$ a sphere, hyperboloid or
flat space.  The ${\cal N}=4^0$ and ${\cal N}=4^-$ theories 
admit similar vacua, but only for $\Omega_2$ a hyperboloid.  

Given the above, we assume an Ansatz of the form $Li_3 \times
\Omega_2$, with any constant curvature internal 2D space.  The metric
is then:
\be
d s^2\,=\,L^2 \left(r^{2z} d t^2-r^2 d x^2-\frac{d r^2}{r^2}\right)
-L^2 d \Omega_2^2 \, ,
\ee
where, $
d \Omega_2^2$ is given by flat space or (\ref{twodgeom}) for the sphere
or hyperboloid, as before.
Similarly, our Ansatz for the gauge fields, motivated by the
symmetries, is:
\bea
{\cal F}_{r t}&=&\frac{\xi_0^2\,\alpha_1}{L} r^{z-1}\,\,;\,\,{\cal F}_{rx}\,=\,
\frac{\xi_0^2\,\beta_1}{L}
\,\,;\,\,{\cal F}_{y_1 y_2}\,=\,\frac{\xi_0^2\,\gamma_1}{L} {\rm e}_2\\
F^{(3)}_{r t}&=&\frac{\xi_0^{-1}\,\alpha_2}{L} r^{z-1}\,\,;\,\, 
F^{(3)}_{rx}\,=\, \frac{\xi_0^{-1}\,\beta_2}{L}
\,\,;\,\,F^{(3)}_{y_1 y_2}\,=\,\frac{\xi_0^{-1}\,\gamma_2}{L} {\rm e}_2 \, ,
\eea
and the scalar field is constant,  $\phi=\phi_0$.  Here, ${\rm e_2}$
is the square-root of the determinant of the metric $d\Omega_2^2$.  Finally, it again
proves useful to rescale the quantities $g_1$ and $g_2$
in such a way as to absorb into their values the free parameters $L$
and $\phi_0$:
\bea
g_1&=& \frac{ \hat g_1 
\xi_0^{-2}}{L}\\
g_2&=& \frac{ \hat g_2 \xi_0}{L} \, .
\eea
Thus in the following we will consider $\hat  g_1$ and $\hat g_2$ as
free parameters, to be fixed by the equations of motion.

As before, this Lifshitz Ansatz reduces the field equations 
to a set of (ten) algebraic equations\footnote{We are grateful to Jerome Gauntlett and Aristomenis Donos for pointing out that Equation (\ref{5dtx}) was missing in the original version of this paper.}:
\bea
\alpha_2 &=&2 \gamma_2 \beta_1+2 \gamma_1
\beta_2\label{5d1} \\
z \beta_2 &=& 2 \alpha_2 \gamma_1+2 \gamma_2 \alpha_1  \label{5d2}\\
\alpha_1 &=& 2 \gamma_2 \beta_2\label{5d3}\\
z \beta_1&=&2 \alpha_2 \gamma_2\label{5d4} \\
0&=&z(z+1)-\frac{4}{3} P-\frac{4}{3} (\alpha_1^2+\alpha_2^2)-\frac23 
(\beta_1^2+\beta_2^2)-\frac23(\gamma_1^2+\gamma_2^2)\label{5d5} \\
0&=&  z+1 -\frac{4}{3} P+\frac23 (\alpha_1^2 + \alpha_2^2)
+\frac{4}{3} (\beta_1^2 + \beta_2^2)-\frac23 (\gamma_1^2+\gamma_2^2) \label{5d7} \\
0&=&z^2+1-\frac{4}{3} P-\frac{4}{3} (\alpha_1^2 + \alpha_2^2)+\frac43
(\beta_1^2 + \beta_2^2)-\frac23
(\gamma_1^2 + \gamma_2^2)
\label{5d6} \\
0&=& \alpha_1 \beta_1 + \alpha_2 \beta_2 \label{5dtx} \\
0&=&-\frac{\lambda}{a^2}-\frac{4}{3} P + \frac23 (\alpha_1^2 +
\alpha_2^2) - \frac23 (\beta_1^2 + \beta_2^2) +\frac{4}{3} 
(\gamma_1^2 + \gamma_2^2)\label{5d8} \\
0&=&\sqrt{\frac32} \frac{d P}{d \phi}+2 (\beta_1^2+\gamma_1^2-\alpha_1^2)-
(\beta_2^2+\gamma_2^2-\alpha_2^2) \, ,\label{5d9} 
\eea
with
\bea
P&=&\frac{\hat g_2}{8}\left( \hat g_2+2\sqrt{2} \hat g_1\right)\\
\frac{d P}{d \phi}&=&\sqrt{\frac23}
\frac{\hat g_2}{4}\left( -\hat g_2+\sqrt{2} \hat g_1\right) \,.
\eea
Here, we have introduced the parameter $\lambda = 1,0,-1$, to include
the cases of the 2D sphere, flat space and hyperboloid,
respectively, at once.

In total we have nine free parameters,  $\alpha_k$, $\beta_k$, $\gamma_k$, 
$a$, $\hat g_1$, $\hat g_2$, with $k=1, 2$,  to fix by means of the
equations.  Although we have more equations than unknowns, as before
it turns out that not all the equations are independent and solutions
can exist for every value of $z \geq 1$. Indeed, it is straightforward
to solve the previous system of
equations completely, 
as we demonstrate in detail in  Appendix \ref{app5dsol}. 
There are
two sets of solutions, which are qualitatively similar:

\smallskip

\noindent
{$\bullet$ $\alpha_1 = 0 = \beta_2$}

\noindent
The solution takes the form
\bea
\alpha_2^2&=&\frac{z(z-1)}{2}\\
\beta_1^2&=&\frac{(z-1)}{2}\\
\gamma_1^2&=&0\\
\gamma_2^2&=&\frac{z}{4}\\
\frac{2\lambda}{a^2}&=&-3z\\
\hat g_2^2&=&2 z^2+3z-2\\
\hat g_1^2&=&\sqrt{2}\left(1+z\right) \, .
\eea
We must have $z\ge 1$.

\smallskip

\noindent
{$\bullet$ $\alpha_2 = 0 = \beta_1$}

\noindent
The solution takes the form
\bea
\alpha_1^2&=&\frac{z(z-1)}{2}\\
\beta_2^2&=&\frac{(z-1)}{2}\\
\gamma_1^2&=&0\\
\gamma_2^2&=&\frac{z}{4}\\
\frac{2\lambda}{a^2}&=&-3z\\
\hat g_2^2&=&-2 z^2+3z+2\\
\hat g_1^2&=&\frac{1}{\sqrt{2}}\left(2z^2 + z + 1\right) \, .
\eea
Here, to ensure that the gauge couplings are real, we must restrict $1
\leq z \leq 2$.

Notice that, since $\lambda < 0 $ in both
cases, the geometry turns out to be restricted to
$Li_3\times H_2$.  Also, similar to the 6D case,
quantization of the 
internal fluxes implies a topological relation between $z$ and the 5D
gauge couplings, $g_1, g_2$.

Although some of the adS$_3 \times \Omega_2$ solutions partially preserve
supersymmetry, breaking ${\cal N}=4$ to ${\cal N}=1$ \cite{romans5d}, it is
again easy to see 
that the Lifshitz fluxes prevent any supersymmetric Lifshitz
configurations.  The spinorial supersymmetry transformations are
(setting $B_{\mu\nu}^{\alpha}=0$):
\bea
\delta \psi_{\mu a} &=& D_{\mu} \epsilon_a + \gamma_{\mu}\left(\frac16
\sqrt{\frac12} {g_2} \xi^{-1} + \frac{1}{12} {g_1} \xi^2
\right) T_{ab} \epsilon^b \nonumber \\
&& \quad - \frac16 \sqrt{\frac16} \left(\gamma_{\mu}^{\,\,\nu\rho} - 4
\delta_{\mu}^{\nu} \gamma^{\rho} \right)\left(\xi F_{\nu\rho}^{(i)} T_{(i)\,ab}
- \sqrt{\frac12} \xi^{-2} {\cal F}_{\nu\rho} \Omega_{ab} \right)\epsilon^b
\\
\delta \chi_a &=& \sqrt{\frac12} \gamma^{\mu} (\partial_{\mu} \phi)
\epsilon_a + \left(\frac12 \sqrt{\frac16} {g_2} \xi^{-1} -
\frac12 \sqrt{\frac13} {g_1} \xi^2 \right) T_{ab} \epsilon^b \nonumber \\
&& \quad - \frac12 \sqrt{\frac16} \gamma^{\mu\nu} \left(\xi
F_{\mu\nu}^{(i)} T_{(i)\,ab} 
- \sqrt{2} \xi^{-2} {\cal F}_{\mu\nu} \Omega_{ab} \right)\epsilon^b
\, ,
\eea
where $T^{(i)}_{ab}, T_{ab}$ are generators of the $SU(2) \times U(1)$
gauge group, and $\Omega_{ab}$ is a metric used to raise and lower any
spinor index.  Plugging our Lifshitz solutions into the dilatino
supersymmetry transformation, we see that three independent projection
conditions would be required to make it vanish, showing that no
supersymmetry can survive.

The 5D Romans' theory has been lifted to  Type IIB supergravity  
in ten dimensions in \cite{lpt} by means
of  an $S^5$ reduction. Building on this result, 11D interpretations
of the 5D theory were given in \cite{5dto11d} and \cite{5dto11dgen}.
Here, we use the 
results of \cite{lpt} to uplift our 5D Lifshitz solutions to solutions
of the Type IIB supergravity equations of motion.

First, we write down the dictionary to go from the conventions in
\cite{lpt} to Romans' conventions used above.  
This requires the following  redefinitions of the fields and 
parameters\footnote{We should also take into account the 
convention $\kappa_5^2 =1/2$ taken in  \cite{lpt}.
Note also that in the final 10D expressions in \cite{lpt} 
they absorbed $g_1,g_2$ in a single $\tilde g$. } (a tilde 
denotes quantities in \cite{lpt} notation): 
\bea
&& \widetilde g_{\mu\nu}  = - g_{\mu\nu} \,, \qquad \qquad \quad \,\,\,\,
\widetilde \phi + 2\widetilde \phi_0 =   2\phi\\  
&& \sqrt{2} \, e^{3\sqrt{2/3}\,\widetilde \phi_0} 
= \frac{g_2}{g_1} \,, \qquad \qquad 
\, \widetilde g = \left(\frac{g_1\,g_2^2}{16} \right)^{1/3}\\  
&& \frac{1}{2} e^{2\sqrt{2/3}\,\widetilde \phi_0} \,
\widetilde { \cal A}_1 = {\cal A}_1\,,  \qquad \qquad   
\frac12 e^{-\sqrt{2/3}\, \widetilde \phi_0}\, \widetilde F^{(i)}_{2}
= F^{(i)}_2\,. 
\eea
Using this dictionary, we can immediately write down our 
ten dimensional solution using the formulae in \cite{lpt}. It is
convenient to define:
\bea 
&& k_0 = \xi_0^{-1} \,\left(\frac{g_2}{g_1\sqrt{2}}\right)^{1/3}  \\
&& C(\rho)= \cos \rho \,, \qquad S= \sin \rho \\
&& \Delta(\rho) = k_0^{-2} \, S^2 + k_0\,C^2  \\ 
&&  U(\rho) = k_0^{-1}\,S^2 +  k_0^2\,C^2 + k_0^{-1} \, ,
\eea
along with the constants:
\bea
&& k_1 = \frac{4 \sqrt{2}}{\xi_0 \, g_1 g_2}, \quad
k_2 = \left (\frac{g_1 \, g_2^2}{2} \right )^{1/3}, \quad 
k_3 = \frac{2\xi_0^{2}}{g_2 k_2}, \quad 
k_4 = -\frac{8k_0^2}{k_2^2 \xi_0^2}, \nonumber \\
&& k_5 = -  \frac{4}{k_2^4}, \quad 
k_6 = -\frac{4\xi_0^{2}}{g_2 k_2^2}, \quad 
k_7 = \frac{\sqrt{2}\,\xi_0k_1}{k_2} , \quad
k_8= - \frac{2}{\xi_0^2 k_0 k_2^3}.
\eea
The solution then reads:
\bea
&&ds^2_{10} = \Delta^{1/2} (Li_3\times d\Omega_2^2) - k_1 \Delta^{-1/2}
\left [\Delta d\rho^2  + k_0 S^2  (d\eta- g_1 {\cal A}_1)^2 
+\frac{C^2}{4k_0^2} \sum_{i}^{3} (h^{(i)})^2 \right ] \nonumber \\
&& {\bf F_5} =  k_2 U \epsilon_5 + k_3 \, C^2 \star_5 F^{(3)}_2 
\wedge \sigma^{(1)}
\wedge \sigma^{(2)} -2 k_3 \,S\,C\, \star_5 F^{(3)}_2 \wedge h^{(3)} 
\wedge d\rho \nonumber \\
&& \hskip9cm + k_4 \star_5 {\cal F}_2 \wedge d\rho \wedge (d\eta - g_1 {\cal
A}_1) \nonumber \\
&&{\bf F_3} = 0, \quad {\bf G_3} = 0, \quad {\Phi} = 0, \quad \chi = 0
\, ,
\eea
and we may also write down the ten dimensional Hodge dual of the
RR five-form as:
\bea
\star {\bf F_5} &=& k_5\,S\,C^3\,U\,\Delta^{-2}d\rho\wedge (d\eta - g_1
{\cal A}_1) \wedge \sigma^{(1)}\wedge \sigma^{(2)}\wedge h^{(3)}
\nonumber \\
&& + k_6 \,S^2\,C^2\,\Delta^{-1} F_2^{(3)} \wedge 
\sigma^{(1)}\wedge\sigma^{(2)}
\wedge (d\eta - g_1 {\cal A}_1) \label{starf5} \\
&& + k_7 \,S\,C \, F_2^{(3)} \wedge h^{(3)}\wedge d\rho \wedge
(d\eta - g_1 {\cal A}_1) + k_8 \, C^4 \, 
\Delta^{-1} {\cal F}_2 \wedge \sigma^{(1)}
\wedge \sigma^{(2)} \wedge h^{(3)} \, .\nonumber
\eea
Here, the 1-forms $h^{(i)}$ are now given in terms
of the left-invariant 1-forms on $S^3$ as:
\be
h^{(i)} = \sigma^{(i)} -  g_2 \,A_1^{(i)} \, ,
\ee
and $\epsilon_5$ is the volume form in the five dimensional
$Li_3 \times \Omega_2$ space. We also recall the 5D
fields ${\cal F}_2$, $\star_5 {\cal F}_2$, $F^{(3)}_2$ and $\star_5
F^{(3)}_2$:
\bea
{\cal F}_2\,&=&\, \frac{\xi_0^2}{L}\,\left[
\alpha_1 r^{z-1}\,d r \wedge dt \,+\, \beta_1 \,d r\wedge d x
\right]\,, \nonumber \\
\star_5 {\cal F}_{2}\, &=&
\, \xi_0^2\,  r^{z} \, {\rm e}_2\,\left[-
\alpha_1 r^{1-z}\,d x \wedge dy_1 \wedge dy_2 \,-\, \beta_1 \,d
t\wedge d y_1 \wedge dy_2 
\right]\,, \nonumber \\
F_{2}^{(3)}\, &=&
\, \frac{\xi_0^{-1}}{L}\,\left[
\alpha_2 r^{z-1}\,d r \wedge dt \,+\, \beta_2 \,d r\wedge d x \,+\, 
\gamma_2\,{\rm e}_2 \,d y_1\wedge d y_2
\right]\,, \nonumber \\
\star_5 F_{2}^{(3)}
\,
&=&
\, \xi_0^{-1}\,   r^{z} \, {\rm e}_2\,\left[-
\alpha_2 r^{1-z}\,d x \wedge dy_1 \wedge dy_2 \,-\, \beta_2 \,d
t\wedge d y_1 \wedge dy_2 \,+\, 
\gamma_2\,{\rm e}_{2}^{-1}\,d t\wedge d x \wedge dr
\right]\,, \nn \\ 
\eea
where $\alpha_k, \beta_k$ ($k=1,2$), $\gamma_2$, $\xi_0$  and $L$ are $z$
dependent constants to be read off from the 5D solution, and
in particular $\alpha_k, \beta_k$ are vanishing when $z=1$.  

As with the previous Lifshitz example, we see the presence of KK gauge 
fields when uplifting the 5D solutions, due to the non-trivial backgrounds 
for $A_1^{(3)}$ and ${\cal A}_1$. However in this case, our
ten dimensional metric is everywhere  regular (apart from the usual
coordinate singularities).  
Again, flux quantization conditions in the
10D system, will lead to further constraints on $z$, $g_1$ and $g_2$.

We can geometrically interpret the ten dimensional
uplifted configuration as follows. When $z=1$, the parameters $\alpha_k$
and $\beta_k$ vanish, so that the 5D ${\cal F}_2$ vanishes and $F_{2}^{(3)}$
has  components only 
in the internal directions. The ten dimensional dual of ${\bf F_5}$,
(\ref{starf5}), is then sourced magnetically by various intersecting D3
branes, and if the D3 brane configuration satisfies certain 
conditions, the system can be supersymmetric.   Meanwhile,
when $z>1$, additional components of ${\bf \star F_5}$ are turned on,
which are sourced both magnetically and electrically by further D3
branes.  The overall
effect is to reduce the symmetry of the three infinite dimensions
from Lorentz to Lifshitz, and to break supersymmetry completely.

\section{Discussion}\label{sectdisc}

In this paper, we provided a simple method that allowed
us to obtain explicit string constructions of Lifshitz 
geometries for general dynamical exponents, $z\ge 1$. Following
a bottom-up approach, our starting point was to look
for Lifshitz solutions in  $d$-dimensional supergravities,
appropriately deforming adS$_{q}\times {\Omega}_{d-q}$ solutions already
known in the literature. Then 
we uplifted them to ten dimensional configurations. 
First, we considered the gauged, massive
${\cal N}=4$ six dimensional supergravity, and showed
that it admits a solution
of the form $Li_4 \times H_2$, with $Li_4$ characterized
by dynamical exponents
$z$ larger than one, 
which may be subject to quantization conditions, and $H_2$ a
hyperboloid that can 
be compact. Then we
discussed the uplifting of this  geometry
to massive Type IIA string theory, giving a basic interpretation
of the resulting configuration in terms of intersecting
branes of various
dimensions. Second, we considered  gauged ${\cal N}=4$ five dimensional
supergravity. 
We found that
this admits solutions of the form $Li_3\times H_2$.
The resulting geometry can be uplifted to IIB string theory,
and can be interpreted as a system of intersecting D3 branes.  
It
would be interesting to study more deeply the brane interpretations of
our 10D configurations.

Our results indeed suggest various issues that deserve further investigation. 
In Ref.\ \cite{gauntlett}, the authors argue that their 10/11D Lifshitz
compactifications can be supersymmetric when they are based on
Sasaki-Einstein manifolds.  In our case in contrast, it is easy to see that
supersymmetry is broken.  For example, looking at the dilatino
supersymmetry transformation in 6D supergravity, the 4D fluxes which
we use to support the Lifshitz geometry lead to a proliferation
of projection conditions.  Since we do not have supersymmetry, there
is no reason to believe that our solutions are stable, and it would be
important to investigate this issue.  Along these lines, it is
intriguing to recall a parallel discussion in the literature on
non-relativistic Schr\"odinger solutions.  Also there, both supersymmetric
and non-supersymmetric solutions have been found in string theory.
Among these are the supersymmetric solutions of \cite{seansch} (albeit
with kinematical supersymmetry only \cite{kinsusy}), which
were surprisingly found to be unstable.  Moreover it was argued in
\cite{seansch} that turning on supersymmetry breaking fluxes can actually
help to restore stability.

We should also note that not all  known
adS$_q\times {\Omega}_{d-q}$ supergravity solutions can be generalized
to Lifshitz solutions. For example,  
${\cal N}=2$ 8D gauged supergravity has
an adS$_4 \times S^4$ background \cite{salamsezgin8d}, but it turns
out that the simple extension to a Lifshitz Ansatz is inconsistent
with the equations of motion.
The same can be said of the adS$_3
\times S^3$ solution to ${\cal N}=4$ gauged, massless, 6D supergravity
\cite{romans6d}.  It would be interesting to understand what
makes our working examples special.  
One characteristic that seems to distinguish them is that the internal
space is a negative curvature hyperboloid, although we do not know yet
whether or not this is a coincidence.   

Lastly, the main advantage of our approach is its simplicity and 
the solutions presented may be useful in developing the Li/CMP correspondence further. 
A possible next step is to find Lifshitz black hole solutions 
and study their properties, which we leave for future work.

\acknowledgments

We are grateful to Sean Hartnoll for valuable discussions and comments
 on the manuscript.  We would also like to thank Cliff Burgess, Ulf
 Danielsson, Jerome Gauntlett, Carlos N\'u\~nez,
 Martin Schvellinger and Thomas Van Riet
  for helpful discussions.   
We acknowledge the Aspen Center for Physics for support
via the Working Group program. 
RG is partially supported by STFC under rolling
grant ST/G000433/1, SLP is supported by the G\"{o}ran
 Gustafsson Foundation, and IZ is partially supported by the
SFB-Transregio TR33 ``The Dark Universe" (DFG) and the EU FP7
program PITN-GA-2009-237920.

\appendix
\section{Appendix: Details of the five dimensional solution }
\label{app5dsol}

In this appendix, we analyse in detail the system of 
the ten equations
(\ref{5d1})$-$(\ref{5d9}) 
associated with Lifshitz configurations in five dimensional 
gauged supergravity. We determine the general solutions
to these equations, providing the values of the nine free
parameters $\alpha_k$, $\beta_k$, $\gamma_k$, $a$, $\hat g_1$, 
$\hat g_2$ as a function of $z$. 

There are combinations of the equations that provide simple relations
among the previous free parameters. 
Taking the differences between (\ref{5d5}) and (\ref{5d6}),
and between (\ref{5d5}) and (\ref{5d7}), we obtain
\bea
\frac{z-1}{2}&=&\beta_1^2+\beta_2^2\label{apps1}\\
\frac{z(z-1)}{2}&=&\alpha_1^2+\alpha_2^2 \, .\label{apps2}
\eea
The first of the previous equations show that $z\ge1$.  
These two equations also imply
\be\label{appdif}
\alpha_1^2-z \beta_1^2 \,=\,z \beta_2^2-\alpha_2^2 \, .
\ee
Meanwhile, multiplying together Eqs. (\ref{5d3}) and (\ref{5d4}), and imposing (\ref{5dtx}), leads to the condition:
\be\label{ab=0}
\alpha_1\beta_1 \,=\,0\,=\,\alpha_2\,\beta_2 \, .
\ee
Thus, we may take either $\alpha_1 = 0 = \beta_2$ or $\alpha_2 = 0 = \beta_1$.  In both cases, Eqs. (\ref{5d1})-(\ref{5d4}) then imply $\gamma_1=0$ and $\gamma_2^2 = z/4$.
Now, combining Eqs.  (\ref{5d7}), (\ref{5d8}) and (\ref{5d9})
one finds
\bea
\frac{\lambda}{a^2}&=&-2\beta_1^2
+2\gamma_2^2-2\beta_2^2-z-1 \label{apps3}\\
\frac{\hat g_2^2}{4}&=&\frac{z+1}{2}+2 \beta_1^2-\alpha_1^2
+\alpha_2^2-\gamma_2^2
\label{apps4}
\\
\frac{\hat g_1 \hat g_2 }{\sqrt{2}}&=&z+1+2 
\beta_2^2+2\alpha_1^2 \, .
\label{apps5}
\eea
There are two sets of solutions, which are qualitatively similar:

\smallskip

\noindent 
$\bullet$  
for $\alpha_1=0=\beta_2$, then $2 \beta_1^2 = z-1$, $2 \alpha_2^2 =
z(z-1)$, $\gamma_1^2 = 0$, $4 \gamma_2^2 = z$, and Eqs. (\ref{apps3})-(\ref{apps5}) imply
\bea
\frac{\lambda}{a^2}&=&-\frac{3}{2}z\\
\hat g_2^2&=& 2 z^2 + 3 z -2\\
\hat g_1^2&=&\sqrt{2}(1+z)
\eea
This solution is valid for all $z \geq 1$.  Notice that the internal
space corresponds to a hyperboloid, since $\lambda$ has to be negative.

\smallskip

\noindent 
$\bullet$  
for $\alpha_2=0=\beta_1$, then $2 \beta_2^2 = z-1$, $2 \alpha_1^2 =
z(z-1)$ , $\gamma_1^2 = 0$, $4 \gamma_2^2 = z$, and Eqs. (\ref{apps3})-(\ref{apps5}) imply
\bea
\frac{\lambda}{a^2}&=&-\frac{3}{2}z\\
\hat g_2^2&=& -2 z^2 + 3 z +2\\
\hat g_1^2&=&\frac{1}{\sqrt{2}}\left(2 z^2 + z + 1\right)
\eea
This solution is physical only for $1 \leq  z \leq 2$.  The
internal space is again a hyperboloid.

\end{document}